\documentclass[aps, 11pt, preprintnumbers,prd, amsmath,amssymb,notitlepage,superscriptaddress]{revtex4} 
\usepackage{bm}
\usepackage{epsfig}
\usepackage{color}
\usepackage{mathtools}
\usepackage{cases}
\usepackage{booktabs}
\usepackage{subcaption}
 \usepackage{url}
\usepackage[
colorlinks=true,
filecolor=black,
anchorcolor=blue,
linkcolor=blue,
citecolor=cyan, 
urlcolor=cyan,
linktocpage=true,
plainpages=false,
breaklinks=true,
pdfstartview=FitH
]{hyperref}

\usepackage[english]{babel}

\usepackage{amsmath,amssymb,graphicx}
\usepackage{caption}
\usepackage{subcaption}
\usepackage{multirow}
\usepackage{dcolumn}
\usepackage{color,url}
\usepackage{colordvi}
\usepackage{bm}
\usepackage{siunitx}
\usepackage{slashed}
\usepackage{natbib}
\usepackage{makecell}
\usepackage{pbox}
\usepackage{array}
\newcommand{\bea}{\begin{eqnarray}}
\newcommand{\eea}{\end{eqnarray}}
\newcommand{\beq}{\begin{eqnarray}}
\newcommand{\eeq}{\end{eqnarray}}

\usepackage{amsmath}
\usepackage{graphicx}

\begin{document}

\title{Gravitational waves from Axion wave production.  }
\author{Mingqiu Li}
\email[]{limingqiu17@mails.ucas.ac.cn}
\affiliation{School of Nuclear Science and Technology, University of Chinese Academy of Sciences, Beijing, P.R.China 100049}
\affiliation{School of Physics Sciences, University of Chinese Academy of Sciences, Beijing 100039, China}
\author{Sichun Sun}
\email[]{sichunssun@gmail.com}
\affiliation{School of Physics, Beijing Institute of Technology, Beijing, 100081, China}
\author{Qi-Shu Yan}
\email[]{yanqishu@ucas.ac.cn}
\affiliation{School of Physics Sciences, University of Chinese Academy of Sciences, Beijing 100039, China}
\affiliation{Center for Future High Energy Physics, Institute of High Energy Physics, Chinese Academy of Sciences, Beijing 100039, China}
\author{Zhijie Zhao}
\email[]{zhijie.zhao@desy.de}
\affiliation{Center for Future High Energy Physics, Institute of High Energy Physics, Chinese Academy of Sciences, Beijing 100039, China}
\affiliation{Deutsches Elektronen-Synchrotron DESY, Notkestr. 85, 22607 Hamburg, Germany}




\begin{abstract}
We consider a scenario with axions/axion-like particles Chern-Simons gravity coupling,  such that gravitational waves can be produced directly from axion wave parametric resonance in the early universe after inflation. This axion gravity term is less constrained compared to the well-searched axion photon coupling and can provide a direct and efficient production channel for gravitational waves. 
Such stochastic gravitational waves can be detected by either space/ground-based gravitational wave detectors or pulsar timing
 arrays for a broad range of axion masses and decay constants.

\end{abstract}
\preprint{DESY-23-137}
\maketitle

\section{ Introduction}

Since the first observation of gravitational waves (GWs), we have acquired a new way to probe the current and early universe and it has a great impact on the fields of astrophysics, cosmology, and even particle physics. GWs at different frequencies from various sources are from the primordial gravitational
waves at $10^{-16}$ Hz\cite{Turner:1996ck,Smith:2005mm,Caprini:2018mtu}, NanoHz $10^{-9}$ Hz\cite{NANOGrav:2023hvm,Li:2023yaj,Kitajima:2023vre,NANOGrav:2023bts,Kitajima:2023cek,Gangopadhyay:2023qjr,Choudhury:2023kam} to the binary system signals up to $10^4$ Hz\cite{Cang:2023ysz}. This whole spectrum of GWs needs different detection schemes, from CMB polarization, Pulsar timing arrays (PTAs)\cite{Manchester_2013,Ferdman_2010,jenet2009north,Hobbs_2010,Manchester_2013_1,Xu:2023wog}, interferometry\cite{Audley:2017drz, Cornish:2018dyw,Luo:2015ght,Guo:2018npi,Gong:2014mca,Coleman:2018ozp,Corbin:2005ny,Musha:2017usi,Punturo:2010zz,Evans:2016mbw,LIGOScientific:2007fwp,LIGOScientific:2014pky,VIRGO:2014yos}, as well as some high-frequency proposals \cite{Aggarwal:2020olq,Gottardi:2007zn,Aguiar:2010kn,Akutsu:2008qv,Nishizawa:2007tn,Ejlli:2019bqj,Cruise_2006,Goryachev:2014yra}.

 Around one-fourth of the universe’s total energy budget is made of dark matter from cosmological and astronomical observations.
Aside from the well-searched weakly interacting massive particles (WIMPs), axion \cite{1983Cosmology,Visinelli:2009kt,Panci:2022wlc,Adams:2022pbo,Co:2017mop,ABBOTT1983133,DINE1983137,Delgado:2023psl} is a promising dark matter candidate, as well as a natural solution to the ‘Strong CP problem’ \cite{Michael1981A,Hook:2018dlk} in quantum chromodynamics (QCD)\cite{Skands:2012ts}. The axion was proposed as the Nambu-Goldstone boson of the spontaneously broken global U(1)PQ symmetry extension of the Standard Model (SM). When the universe cools down to the QCD scale the axion acquires a tiny mass and becomes a pseudo-Nambu-Goldstone particle. 
 After the QCD phase transition, the axion field begins to oscillate and the energy density from classical oscillation can play the role of dark matter\cite{Co:2017mop}.  
In well-motivated scenarios such as the Dine-Fischler-Srednicki-Zhitnitsky (DFSZ) \cite{PhysRevD.107.095020,Zhitnitskij1980OnPS,Michael1981A} and the
Kim-Shifman-Vainshtein-Zakharov (KSVZ) \cite{PhysRevLett.43.103,SHIFMAN1980493} models, QCD axions's characteristic axion-gluon
coupling is generated via  SU(3)c-charged fermion loops. 
In general axion-like particles (ALPs) are light bosons
that have electromagnetic Chern-Simons (CS) type of coupling but are not necessarily coupled to gluons. ALPs can also acquire axion-graviton Chern-Simons coupling,  which can be generated through heavy particle loops.  The axion-graviton coupling can also induce axion photon coupling with Planck suppression \cite{Sun:2020gem}. ALPs are abundant
in string-theory motivated models \cite{Svrcek:2006yi,Marchesano:2014mla}; ALPs generally
can have a much wider mass range as the dark
matter candidate. We will use ‘axion’ to denote both the
QCD axion and ALPs.

The gravitational wave from axion was considered with the axion dark gauge boson coupling, to avoid the stringent constraints from axion photon couplings. For the inflationary scenario that axion is the inflaton, axion gauge boson coupling can induce large primordial GWs e.g. in the so-called axion monodromy model through tachyonic instability \cite{McAllister:2008hb,DiLuzio:2020wdo,Hebecker:2016vbl,Anber:2009ua,Barnaby:2012xt, Anber:2012du, Domcke:2016bkh}. If considering the mechanism that axion oscillates after inflation, tachyonic instability in a dark gauge field induced by an axion-like particle is a known source of dark matter and/or stochastic GWs \cite{Madge:2021abk,Machado:2018nqk,Machado:2019xuc,Kitajima:2020rpm,Geller:2023shn, Dror:2018pdh,Co:2018lka,Bastero-Gil:2018uel,Agrawal:2018vin}, and friction for relaxion\cite{Hook:2016mqo,Fonseca:2018xzp}.

Here in this paper, we consider a less studied axion graviton Chern-Simons coupling \cite{Alexander:2009tp,Dyda:2012rj} in the early universe after inflation, such that this term can produce GWs directly through axion rolling parametric resonance. The effect of such coupling term has been studied in inflationary scenarios for primordial gravitational wave production, and binary merger of black holes \cite{Alexander:2009tp}. However, for high-scale inflationary scenarios, the ghost issues in Chern-Simons gravity may become severe. In this work, the frequency of directly produced GWs is $10^{-9}\mathrm{Hz}\sim 10^{-2}\mathrm{Hz}$.  Due to the large parameter space of axion mass, we can observe such gravitational wave spectrum as low as PTA band\cite{Manchester_2013,Ferdman_2010,jenet2009north,Hobbs_2010,Manchester_2013_1} when $m_a \sim 10^{-12}$eV, up to space interferometry 0.01 Hz \cite{Audley:2017drz,Cornish:2018dyw,Luo:2015ght,Guo:2018npi} when $m_a \sim 1$eV. The gravitational wave spectrum here has a peak frequency and is spread out around one order of magnitude by the universe expansion.

This paper is organized as follows. In section \ref{s2} we briefly review the axion evolution equations in the early universe with axion graviton Chern-Simons coupling and universe expansion. We study how the axions produce GWs in section \ref{s3}. In section \ref{sghost}, we address the issue how other regularization procedures by introducing high order terms can affect our results. In section \ref{s4}, we summarize and discuss our result. 

\section{ The Model}
\label{s2}
We begin with the  action of Chern-Simons modified gravity with axion coupling:
\begin{equation}
\begin{aligned}
 S=\int d^4x\sqrt{-g}\left(\frac{ R}{16\pi G}+\frac{\alpha}{4}\phi R\Tilde{R}+\frac{1}{2}\partial_\mu \phi\partial^\mu \phi-V(\phi)\right)   
\end{aligned}    
\end{equation}
where $R$ is the Ricci scalar, $ R\Tilde{R}=\frac{1}{2}\epsilon^{\rho\sigma\alpha\beta}R^{\mu\nu}_{\alpha\beta}R_{\nu\mu\rho\sigma}$, G is the gravitational constant, and the potential of axion field is
\begin{equation}
    V(\phi)=m_a^2f_a^2\left[1-\cos\left(\frac{\phi}{f_a}\right)\right],
\end{equation}
where $m_a$ is the mass of axions and $f_a$ is the decay constant. 
We will consider the linear expansion of GW field under the background of Friedmann-Robertson-Walker metric, which can be expressed as 
\begin{equation}
    ds^2=-dt^2+a^2(t) \delta_{ij} dx^idx^j\,,
\end{equation}
where the $a(t)$ is the cosmic scale factor. Neglecting the backreaction of GWs, the  equation of motion (EOM) of the axion field can be expressed as  \cite{Machado:2018nqk}
\begin{equation}
\label{eqphi}
    \ddot{\frac{\phi}{f_a}}+3H\dot{\frac{\phi}{f_a}}+m_a^2\sin \frac{\phi}{f_a}=0
\end{equation}
where $H$ is defined as $H=\dot{a}(t)/a(t)$. In the epoch that the Hubble parameter $H\gg m_a$, $\phi$ is almost constant. When the universe cools down to $H\approx m_a$, the axion field begins rolling toward the minimum of the potential and then oscillates. Assuming a radiation-dominated universe, the temperature when $H= m_a$ is
\begin{equation}
\begin{aligned}
\label{eqT}
    T_*&=\left(\frac{45}{4\pi^3}\right)^{1/4}g_{*}^{-1/4}m_a^{1/2}G^{-1/4}\\
    &=2.7118\times 10^4 \left(\frac{g_*}{100}\right)^{-1/4}\left(\frac{m_a}{\mathrm{eV}}\right)^{1/2}\mathrm{GeV}
\end{aligned}
\end{equation}
for a wide range of axion mass, i.e., $m_a\in[10^{-12}\mathrm{eV},1\mathrm{eV}]$, the radiation dominated assumption is consistent \cite{Machado:2018nqk}. In the following study, to solve Eq.(\ref{eqphi}), 
we chose initial conditions $\phi(t_*)/f_a=\theta\sim \mathcal{O}(1), \dot{\phi}(t_*)=0, H(t_*)=(2t_*)^{-1}=m_a$.

When $t\gg t_*$ we have $\phi/f_a\ll 1$. Thus the solution of Eq. (\ref{eqphi}) can be put as
\begin{equation}
	\frac{\phi}{f_a}=\frac{\pi \theta}{4(2m_at)^{1/4}}\left[J_{5/4}\left(\frac{1}{2}\right)Y_{1/4}(m_at)-Y_{5/4}\left(\frac{1}{2}\right)J_{1/4}(m_at)\right],
\end{equation}
where $J_n(z),Y_n(z)$ are Bessel functions of the first and second kind.
When $z=m_at\gg 1$, using $a(t)=\sqrt{2 m_a t}, J_n(z)\simeq \sqrt{\frac{2}{\pi z}}\cos\left(z-\frac{(2n+1)\pi}{4}\right), Y_n(z)\simeq \sqrt{\frac{2}{\pi z}}\sin\left(z-\frac{(2n+1)\pi}{4}\right)$, $\dot{\phi}$ can be cast into the form 
\begin{equation}\label{eqphi001}
    \frac{\dot{\phi}}{f_a}\simeq \phi_0 m_a a(t)^{-3/2}\sin (m_at+\psi_0).
\end{equation}
where $\phi_0$ can be expressed as 
\begin{equation}\label{eqphi002}
    \phi_0 =\frac{\sqrt{\pi}\theta}{2}\sqrt{J_{5/4}\left(\frac{1}{2}\right)^2+Y_{5/4}\left(\frac{1}{2}\right)^2}\simeq 1.66.
\end{equation}

To derive the linear equation of motion of GWs, we consider the perturbed metric with the following form
\begin{equation}
    ds^2=-dt^2+a^2(t)\left(\delta_{ij}+h_{ij}(t,\mathbf{x})\right)dx^idx^j,
\end{equation}
where the cosmic scale factor $a(t)$ can be taken as $a=\sqrt{t/t_*}=\sqrt{2tm_a}$ for radiation dominated universe as mentioned above. Notice we have chosen $a(t_*)=1$ in the whole study.
The GW fields $h_{ij}(t,x)$ can be expanded as $h_A(t,k)$ in the momentum space by the following Fourier transformation 
\begin{equation}
  h_{ij}(t,\mathbf{x}) =\sum_{A=R,L}\int \frac{d^3k}{(2\pi)^3}h_A(t,k)e^{i\vec{k}\cdot\vec{x}}e_{ij}^A
\end{equation}
where $A =L, R$ labels the left-handed and right-handed polarization, respectively.

Thus, the quadratic action for the gravitational waves from our model reads
\begin{equation}
\begin{aligned}
S_h&=\frac{1}{32\pi G}\sum_{A={\mathrm{R,L}}}\int d t\int\frac{d^{3}k}{(2\pi)^{3}}\,a^{3}D_{A}\left(\dot{h}_{A}{}^{2}-\frac{k^{2}}{a^{2}}h_{A}{}^{2}\right) \label{quadS}
\end{aligned}	
\end{equation}
where $\quad D_A=1-\frac{\lambda_A k}{am_{cs}}$. In our convention,  $\lambda_R=1$ and $\lambda_L=-1$. The Chern-Simons mass is defined as 
\begin{equation}
    m_{cs}(t) =\frac{1}{16\pi G\alpha\dot{\phi}(t)}\,,
\end{equation}
We point out that Chern-Simons mass $m_{cs}(t)$ is a time-dependent parameter. Furthermore, it can be either positive or negative since $\dot{\phi}$ can be either positive or negative. This time-dependent definition is the same with the literature as Ref.\cite{Alexander:2009tp,Dyda:2012rj}. The EOM for $h_A$ can be derived from this action \cite{Chu:2020iil},
\begin{equation}
	\begin{aligned}
		\ddot{h}_A+ 2 Q_A \dot{h}_A+\frac{k^2}{a^2}h_A & =0\,,
		\label{eqhA}
	\end{aligned}
\end{equation}
where $Q_A = \frac{1}{2} \left(3H+\frac{\dot{D}_A}{D_A}\right)$.

To examine the tachyonic GWs production, from Eq. (\ref{eqhA}), we can replace the graviton field $h_A$ by a new field $\psi_A$ which is defined by the relation 
$h_A  =  \psi_A \exp\left [  - \int^t Q_A(t^\prime)  dt^\prime \right ]$, thus we arrive at the following the equation of GW field $\psi_A$
\begin{equation}\label{eqpsia}
\ddot{\psi}_A + \left [ \frac{{k}^2}{a^2} - \dot{Q}_A(t) -  Q_A(t)^2\right ] \psi_A=0 \,.
\end{equation}
The tachyonic instability can happen when the following condition is satisfied
\begin{equation}
\omega^2 = \frac{{k}^2}{ a^2} - \dot{Q}_A(t) - Q_A(t)^2 < 0\,,\label{omega2}
\end{equation}
which can lead to an exponential enhancement of GWs.

Besides the tachyonic instability, the EOMs of GWs also have included narrow parameter resonances. To examine where and when the narrow parameter resonances might occur, we can 
define the parameter $\delta=16\pi G\alpha f_am_a^2\phi_0/a^{3/2}=m_a/\text{min}(|m_{cs}|)$. In the limit that $\delta\ll 1$, we can substitute the result of Eq.(\ref{eqphi001}) into $\omega^2$ defined in Eq. (\ref{omega2}), and we arrive at the following result
\begin{equation}
\begin{aligned}
  \omega^2&=\frac{k^2}{a^2}+\frac{3}{4}H^2-\frac{\lambda_Akm_a\delta}{2a}\cos(m_at)\\
  &+ \frac{\lambda_Akm_a\delta}{a}\sin(m_at)\frac{H}{m_a} +\frac{15\lambda_Akm_a\delta}{8a}\cos(m_at)\frac{H^2}{m_a^2}+\mathcal{O}(\delta^2)\\
  &\simeq \frac{k^2}{a^2}-\frac{\lambda_Akm_a\delta}{2a}\cos(m_at).
\end{aligned}    
\end{equation}
Now $  \omega^2<0$ leads to $k/a <\frac{m_a^2}{2\text{min}(|m_{cs}|)} $. 
To obtain these results, we have used the condition $t\gg t_*$ and $H\ll m_a$.
Finally, the EOM given in Eq.(\ref{eqpsia}) can be cast into the well-known Mathieu equation\cite{brimacombe2021computation},
\begin{equation}
\begin{aligned}
&\frac{d^2\psi_A}{dx^2}+\left(A_k-2q\cos(2x)\right)\psi_A=0\\
\mathrm{with}\quad & x=m_at/2,\quad A_k=\frac{4k^2}{a^2m_a^2}, \quad q=\frac{\lambda_Ak\delta}{m_aa}.
\end{aligned}    
\end{equation}
As is well known, the narrow parameter resonances can occur when $A_k=1$. Therefore, it should be pointed out that both narrow resonances and tachyonic instability can lead to an exponential enhancement of GWs. In this work, we will mostly deal with the narrow parameter resonances.

It is necessary to point out that there are three fixed energy scales in our model,i.e. $\Lambda_{Planck}$, $f_a$, and $m_a$. The Planck energy scale $\Lambda_{Planck}$ introduces a natural momentum cutoff for $k$, above this scale, the CS gravity should be embedded into more fundamental theories like the string theory or quantum gravity theory. Naturally, we have $\Lambda_{Planck} > f_a > m_a$. While there are two time-dependent energy scales, i.e. $H$ and $|m_{cs}(t)|$, the Chern-Simons mass $|m_{cs}(t)|$ will in general be a function of time or redshift. as widely used in the literature. Before the epoch $H(t^*)=m_a$, the function $\dot{\phi}(t^*) \approx 0$ and leads to the result $|m_{cs}(t^*)| \approx \infty$, and general relativity is recovered, as argued in ref. \cite{Dyda:2012rj}.

There are two major issues on the action given in Eq. (\ref{quadS}) and the EOM given in Eq. (\ref{eqhA}). 1) There exist ghost modes when $k$ is too large and make the factor $D_A$ negative. Ghost modes are negative energy modes, which lead to an unbounded-below Hamiltonian where the vacuum state is not stable \cite{Dyda:2012rj}. Since $\dot{\phi}(t)$ can be either positive or negative, ghost modes can appear in either right or left polarizations. 2) The EOM of GWs encounters singularities when $D_A \approx 0 $, as considered in the context of inflation by \cite{Alexander:2004wk,Alexander:2007kv}, which indicates the breakdown of linear expansion of the action and the numerical solutions are undetermined and unphysical.

To remove ghost modes, conventionally, a hard cutoff $\Lambda$ in the momentum space $k$ can be introduced and only modes $k< \Lambda$ are considered. For example, in the weak coupling limit $\alpha f_a \ll 1$, the cutoff $\Lambda$ in the momentum space is close to the Planck scale $\Lambda_{Planck}$. While in the strong coupling limit $\alpha f_a\gg 1$, the cutoff $\Lambda$ can be much smaller than the Planck scale and can be close to $m_a$. Since momentum modes $k>\Lambda$ could be produced by the early universe processes like inflation or phase transitions before the epoch $H(t^*)=m_a$, it is not justified to simply neglect them. 

To cure these two issues and to obtain a reasonable result, it is necessary to know where the ghost modes are situated in the phase space of $k$ and $t$. For numerical analysis, we should choose some benchmark points in the parameter space of the model. There are 4 free parameters, $\alpha$, $f_a$, $m_a$, and $\theta$. We set $\theta=1$ and choose five benchmark points presented in Table \ref{table1}, which are consistent with all experimental bounds. 
For BP1, BP2 and BP3 in table \ref{table1}, it is found that $\delta\simeq \left(\frac{9.2}{a}\right)^{3/2}$ and the condition $\delta <1$ can be fulfilled when $a>10$ ($t m_a > 50$ or so). For BP4 and BP5, it is found that the condition $\delta\simeq \left(\frac{5.8}{a}\right)^{3/2}<1$ can be fulfilled when $a>6$ ($t m_a > 20 $).
\begin{table}[h]
\setlength{\tabcolsep}{12pt}
\begin{center}   
\label{table:1} 
\begin{tabular}{|c|c|c|c|}   
\hline    
\hline   
--& $m_a  $ (eV) & $f_a$ (GeV) & $\alpha$ ($\mathrm{GeV}^{-1}$)  \\ 
\hline 
  BP1(ALP)& $1$  &  $ 10^{17}$   &   $5\times 10^{38}$\\
\hline
BP2(ALP)& $10^{-12}$  &  $ 10^{17}$   &   $5\times 10^{62}$\\
\hline
BP3(QCD axion)& $2\times 10^{-12}$  &  $ 3\times 10^{18}$   &   $4.2\times 10^{60}$\\
\hline
BP4(QCD axion)& $2\times 10^{-12}$  &  $3\times 10^{18}$   &   $2.1\times 10^{60}$\\
\hline 
  BP5(ALP)& $1$  &  $ 10^{16}$   &   $2.5\times 10^{39}$\\
\hline
\hline   
\end{tabular}
\caption{Parameters for 5 benchmark point are presented, we have fixed $\theta=1$, where BP3 and BP4 are consistent with the QCD axion. The unit of each parameter is also provided.}
\label{table1}
\end{center}   
\end{table}

\begin{figure}[t]
	\centering
	\includegraphics[width=0.7\linewidth]{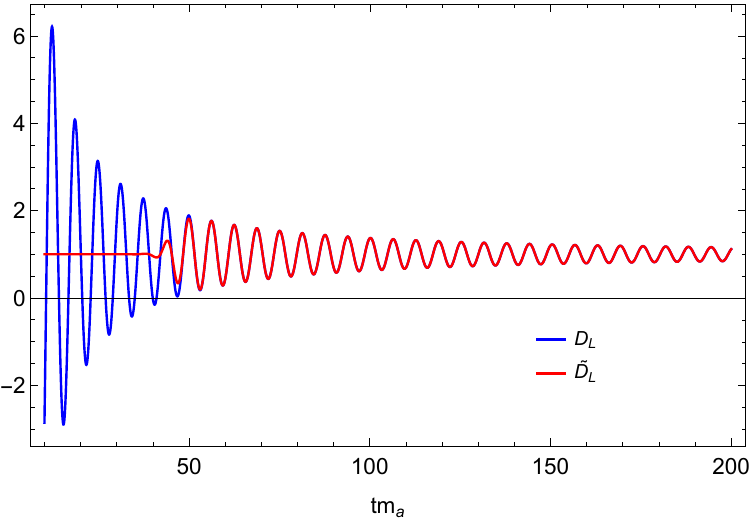}
	\caption{ Blue line is $D_L(t)$ and $\tilde{D}_L=1+\frac{k}{a |m_{cs}|}\Theta\left(1-\frac{k}{a\Lambda}\right)$ is showed in red. We use the parameters of BP1 given in Table \ref{table1} and  $k/m_a=10$ as input.}
	\label{fig:plot5}
\end{figure}

As an example, we show  $D_L$ in Fig.\ref{fig:plot5}. To conduct numerical solutions, we have introduced dimensionless parameters $t m_a$ to replace $t$  and $k/m_a$ to replace $k$. There are two comments on the behavior of $D_A$.
\begin{itemize}
    \item  When $t m_a >50$, $D_L>0$ consistently and we can safely study how the $\phi$ enhance $h_L$ using Eqs.(\ref{eqhA}). 
    \item When $t m_a <50$, $D_L$ can be negative, and the ghost modes can appear. We aim to study the resonant amplification of GW normal mode at $k/a(t)\sim m_a/2$. 
\end{itemize}
It should be emphasized that ghost mode does not affect the resonant amplification at $t m_a >50$. 

Therefore, to have a result without the effects of ghost modes and the dependency upon extra free parameters, we introduce a time-dependent parameter $\Lambda < |m_{cs}|$ to regularize the effects of ghost modes and to avoid the singularities in the EOM of GWs. At any a time $t$, the cutoff $\Lambda$ can be found by imposing the condition $k/(a |m_{cs}|) <  k/(a\Lambda)<1$. Then for any momentum mode $k$ which satisfies $k< \Lambda$, we can use Eq.(\ref{eqhA}) to obtain the corresponding GW solutions. 

In practice, at any time $t$, we choose $\Lambda$ to be slightly smaller than the amplitude of the oscillating $|m_{cs}|$. 
Our theory is now parameterized by two parameters, the cutoff $\Lambda$, and the Chern-Simons mass $|m_{cs}|$. Now since
the ghost modes arise only for momentum k satisfying $k/a > |m_{cs}|$, thus in the region with $k/a < \Lambda < |m_{cs}|$, there is no ghost modes anymore.


In other words, our regularization method is equivalent to turning on the CS term when $k/a<\Lambda$. 
 This can be realized by making replacement
 \begin{equation}
 	\frac{k}{a m_{cs}}\longrightarrow \frac{k}{a m_{cs}}\Theta\left(1-\frac{k}{a\Lambda}\right), \mathrm{with}\quad 
 	\Theta(x)=\begin{cases}
 		1, & x>0,\\
 		0, & x \le0.
 	\end{cases}    
 \end{equation}
 When $|m_{cs}|>\Lambda>k/a$, $\Theta\left(1-\frac{k}{a\Lambda}\right)=1, D_A>0$, the equations stay unchanged. When $\Lambda<k/a$, $\Theta\left(1-\frac{k}{a\Lambda}\right)=0$, CS term is turned off to avoid $D_A<0$. 
 We  define $\tilde{D}_L= 1+\frac{k}{a m_{cs}}\Theta\left(1-\frac{k}{a\Lambda}\right)$.
 Fig.\ref{fig:plot5} shows how the replacement works. When $t m_a>50$, the theory stays the same after regulation. When $t m_a <50$, the $D_L$ is  replaced by one and the $\dot{D}_L$ term in Eqs.(\ref{eqhA}) vanish. 

In our numerical analysis, since in Eq.(\ref{eqphi001}) the amplitude of $\dot{\phi}$ is proportional to $f_a m_a a(t)^{-3/2}$, we choose $\Lambda=\left(16\pi \,G\, \alpha a(t)^{-3/2}\theta f_a\left(A_1 m_a+\frac{A_2}{t}\right)\right)^{-1}$, where $A_1=1.66, A_2=-0.12$ are numerically determined by requiring the cut-off scale to be less than the oscillating $|m_{cs}|$. 

By introducing such a regularization procedure, we can remove ghost modes and avoid the singularities in the EOM of GWs. We can also separate the ghost regions and parameter resonance regions. In the meantime, we can guarantee the validity of the ansatz neglecting the backreaction of GWs to the axion field. We will address the issue of how our results might be affected by other regularization procedures including higher order terms in the later section.

 In Ref.\cite{Dyda:2012rj}, a cutoff scale $\Lambda_c$ on the physical wavenumber   $k/a$ is introduced to avoid an infinite vacuum decay rate. For $k/a<|m_{cs}|$, there is no ghost mode. For $\Lambda_c>k/a>|m_{cs}|$, decay of the vacuum is allowed but the decay rate is constrained by observational bounds. In this work, we focus on $k/a<\Lambda \le |m_{cs}|$, without running into the vacuum decay problem.

\begin{figure}[htb]
\centering
\includegraphics[width=0.7\linewidth]{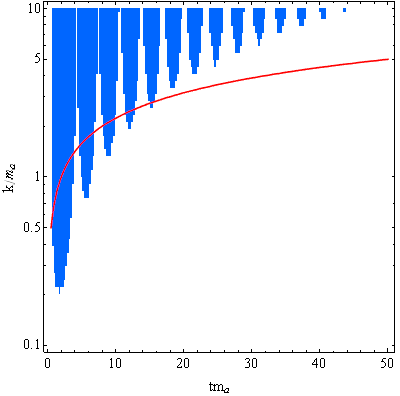}
\caption{The regions of ghost modes (the blue regions) and the resonance region (the red line) are shown, where the model parameters of BP1 are used as input.}
\label{plotDA}
\end{figure}

In Fig.\ref{plotDA}, the phase space of ghost modes are shown by the blue regions corresponding to $k/a>|m_{cs}|$. It is noteworthy that the regions of ghost modes oscillate with time since $|m_{cs}|$ changes with axion field $\dot{\phi}$. 
The parameter resonance region is depicted by a red curve. When $t m_a > 50$, the ghost modes regions and parameter resonance regions are separated from each other, i.e. there are no overlapping areas between these two types of regions. Our regularization procedure is adopted to remove the effects of ghost modes, i.e. the contributions of strong coupling regions. 

\section{Resonant amplification of GWs}
\label{s3}
The enhancement of GWs passing through the axion cloud has been studied in \cite{Sun:2020gem, Yoshida:2017cjl, Chu:2020iil, Lambiase:2022ucu}.  Those studies mainly focus on a much later epoch when $H\ll m_a$ and the universe expansion effects can be neglected. Their result shows that a resonant amplification of GWs happens when $k=m_a/2$. 

\begin{figure}[htb]
\centering
\captionbox{
The evolutions of $h_A$ are shown for the momentum mode $k=10m_a$. We have used BP1 as input for the left plot, and BP5 for the right plot.
\label{plot2}
}{ 
\includegraphics[width=0.48\linewidth]{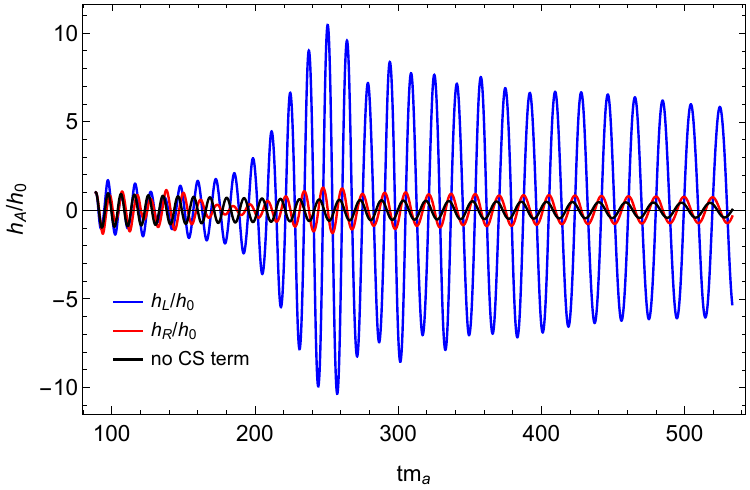}
\includegraphics[width=0.48\linewidth]{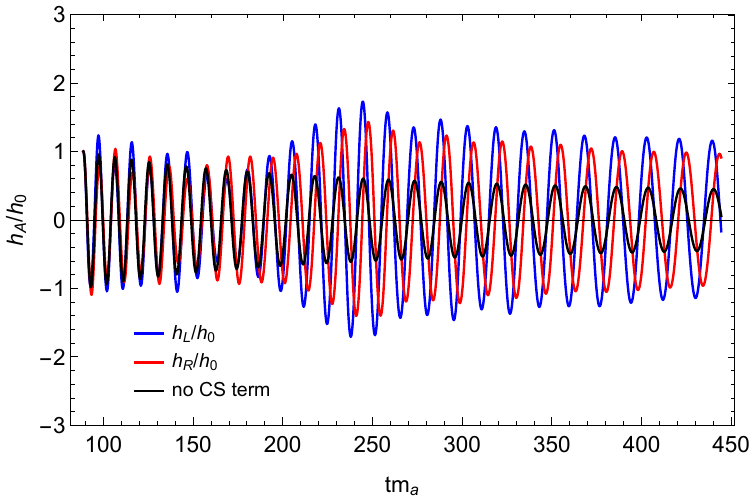}
}
\end{figure}

However, when the universe's expansion being considered, it becomes different. Fig.\ref{plot2} shows how the amplitudes of GWs evolve for the mode $k=10m_a$, where $h_A/h_0$ is the amplitude of GWs normalized by the initial value. 
\begin{itemize}
    \item In the case $k/a\gg m_a$, the amplitudes are almost unchanged. 
    \item As time goes by, and the value of $k/a$ decreased. When the condition $k/a\sim m_a/2$ is met, resonant amplification occurs, and $h_L$ increases exponentially. It is observed that right helicity mode $A=R$ and left helicity mode $A=L$ evolve dramatically differently, which can be attributed to the fact that the CS-modified gravity breaks the parity symmetry. In the transformation $\phi\rightarrow -\phi, L\leftrightarrow R $, the Eq.\ref{eqhA} is unchanged, which means that these two polarization modes behaviors can switch if we use a different initial conditions $\phi(t_*)=-f_a\theta$. 
\item When $a$ becomes sufficiently large, $k/a$ becomes much smaller than $m_a/2$, resonant amplification stops and $h_A$ decreases slowly as the universe expands.
\end{itemize}

It is observed that, as shown in Fig.\ref{plot2}, resonant amplification of GWs happens at $t m_a \sim 200$ for our benchmarks BP1 snd BP5 given in Table \ref{table1}.

\begin{figure}[htb]
\centering
\includegraphics[width=0.7\linewidth]{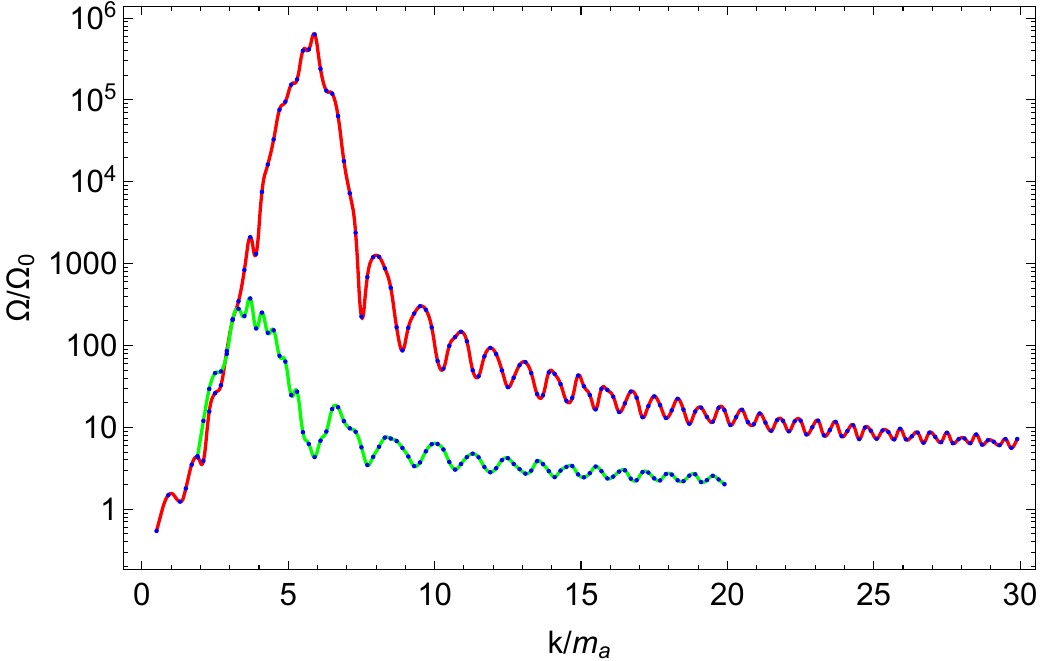}
\caption{Magnification of energy density for GWs are shown, where the red line denotes the case of BP1, while the green line denotes the case of BP5.  }
\label{plot3}
\end{figure}

For different values $k/m_a$, we numerically solve Eq.(\ref{eqhA}) and obtain magnification of GWs. The energy density $\Omega \propto h_A^2$. When $a\gg 2k/m_a$, $\Omega/\Omega_0$ trends to a constant, where $\Omega_0$ is the energy density when  CS modification is absent. The results are shown in Fig.\ref{plot3}, where the red curve is for the case of BP1 and the green curve is for the case of BP5. 

As an example, we highlight some salient features of the BP1 denoted by the red curve. 
\begin{itemize}
    \item For the momentum modes $k/m_a \gg 10 $, to meet the resonance condition $k/a=m_a/2$, a much longer time for the cosmic scale $a(t)$ to evolve is demanded. A longer time means a weaker axion field, which decreases with the increase of time, thus a smaller magnification is expected. 

\item For the momentum modes $k/m_a <5$, to meet the resonance condition $k/a=m_a/2$, a much shorter time for the cosmic scale $a(t)$ to evolve is required. Thus even when resonant amplification happens, the momentum modes $k/a$ develop too fast to allow the resonant amplification to grow sufficiently. 

\item It is observed that the peak of GWs is at $k/m_a \sim 6$ and the magnification factor can reach $10^6$ or so.

\end{itemize}
It should be mentioned that for the case of BP5, as Fig.\ref{plotDA} shows, for the momentum modes $k/m_a <3 $, ghost modes are encountered due to the condition $\left|\frac{k}{a m_{cs}}\right|>1$. Even though the resonant enhancement can occur, to remove ghost modes, we have to turn on our regulator.

To make meaningful predictions and to solve the EOM of GWs, we must introduce some initial conditions for the GWs. GWs of inflation can be our initial conditions. Inflation can generate a scale-invariant gravitational wave spectrum. The spectrum can be approximated by\cite{Saikawa:2018rcs} 
\begin{equation}
\begin{aligned}
    \Omega_{\mathrm{gw}}h^2(\tau_{0},k)&\approx\frac{1}{24}\Omega_{\gamma}h^2\left(\frac{g_{\star\rho,\mathrm{hc}}}{2}\right)\left(\frac{g_{\star s,\mathrm{hc}}}{g_{\star s,\mathrm{fin}}}\right)^{-\frac{4}{3}}\mathcal{P}_{T}(k)\\
    &\approx1.29\times10^{-17}\left(\frac{g_{*s,\mathrm{fin}}}{3.931}\right)^{\frac{4}{3}}\left(\frac{g_{*s,\mathrm{hc}}}{106.75}\right)\left(\frac{g_{*s,\mathrm{hc}}}{106.75}\right)^{-\frac{4}{3}}\left(\frac{V_{\mathrm{inf}}^{1/4}}{10^{16}\,\mathrm{GeV}}\right)^{4},
\end{aligned}
\end{equation}
where $V_{\mathrm{inf}}^{1/4}$ is the energy scale of inflation. If there exist axions, the GWs from inflation can be magnified at certain specific frequencies. We can also consider the other sources as the seeds for the stochastic GWs for our axion enhancement study. Here for the sake of simplicity, we just take the simplest scale invariant spectrum from inflation as an example. Suppose GWs induced by inflation is about $h^2\Omega_0\sim 10^{-16}$\cite{Saikawa:2018rcs}, \textcolor{black}{which is well below the upper bound on the amount of radiation $\int  \Omega_{GW}(k)d\ln k\leq 5.6\times 10^{-6}\Delta N_{eff}$ \cite{Aggarwal:2020olq}, where $\Delta N_{eff}$ is the extra neutrino species and roughly satisfes the condition $\Delta N_{eff}<0.2$. } 

\begin{figure}[htb]
\includegraphics[width=0.7\linewidth]{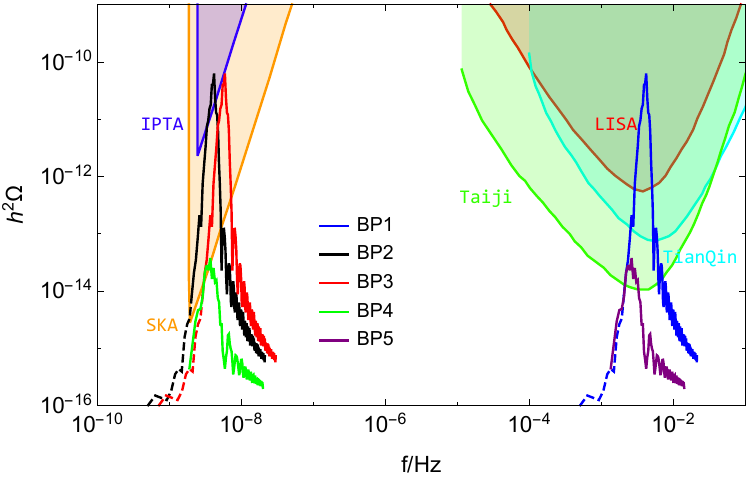}
\caption{ GWs spectra after the amplification by axions are shown. The dashed part of each line is described in the text. We plot 5 benchmark points as given in Table \ref{table1}. The sensitivity curves from Taiji\cite{Guo:2018npi}, TianQin\cite{Luo:2015ght}, LISA\cite{Audley:2017drz} as well as IPTA\cite{Hobbs_2010}, SKA\cite{5136190, Moore:2014lga}, are shown. }
\label{plot4}
\end{figure}

With this initial condition, we show the energy density after the amplification by axions in Fig.\ref{plot4}. The energy density shape depends on the magnification of Fig.\ref{plot3} 
and the peak frequencies are given as 
 \begin{equation}
  \begin{aligned}
  f&=\frac{k}{2\pi}\frac{a_*}{a_0}=\frac{k}{2\pi}\frac{T_0}{T_*}\left(\frac{g_0}{g_*}\right)^{1/3}\\
 &\approx7.125\times 10^{-4}\left(\frac{100}{g_*}\right)^{1/12}\left(\frac{k}{m_a}\right)\left(\frac{m_a}{\mathrm{eV}}\right)^{1/2} \mathrm{Hz}.    
  \end{aligned}   
 \end{equation}
 
To plot the results, we have used Eq.(\ref{eqT}) and the parameters $T_0=2.725\mathrm{K}$ and $g_0=3.938$. The value $k/m_a$ can be directly taken from Fig.\ref{plot3}. The model parameters for each of the benchmark points shown in Fig.\ref{plot4} are presented in Table \ref{table1}. In Fig.\ref{plot4}, the dashed part of each line denotes the ghost mode region $\left|\frac{k}{a m_{cs} }\right|>1$ even when the resonance enhancement condition $k/a=m_a/2$ is met. To regularize these ghost modes, we use our regulator. 

As shown by this figure, after the axion enhancement, the GWs can be detected by future GW detectors, like Taiji\cite{Guo:2018npi}, TianQin\cite{Luo:2015ght}, LISA\cite{Audley:2017drz} for axion mass $m_a=10^{-3}\mathrm{eV}\sim 1\mathrm{eV}$ and by IPTA\cite{Hobbs_2010}, SKA\cite{5136190, Moore:2014lga} for $m_a=10^{-12}\mathrm{eV}$, respectively.

One may worry whether our model parameters are consistent with the ansatz neglecting the backreaction of GWs. It is useful to examine the values of $\Omega_{\phi}$ and $\Omega_{gw}$ at the epoch $H(t^*)=m_a$ and later. The frame-dragging experiments gives a limit for CS gravity  coupling by\cite{Smith:2007jm,Alexander:2009tp} $\kappa^{-1} \alpha\dot{\phi}<3000 \mathrm{km}$, or $\alpha\Omega_{\phi}^{1/2}<5.25\times 10^{81}\mathrm{GeV}^{-1}$, where $\Omega_{\phi}$ is energy fractions of axions nowadays.  
The energy fraction of axions at the epoch $H(t^*)=m_a$ can be found as $\Omega_{\phi*}\sim 2.6\times 10^{-4}\left(\frac{f_a}{10^{17}\mathrm{GeV}}\right)^2 $, which is consistent with the radiation dominated universe. Energy fractions of GWs at the epoch $H(t^*)=m_a$ is found to be $\Omega_{gw*}\simeq 2.78\times 10^{4}\left(\frac{g_*}{100}\right)^{1/3}\Omega_{gw}(T=2.7K)$. Using its today's value $\Omega_{gw}(T=2.7K)\sim 10^{-10}$, we can found that the energy density $\Omega_{gw*}\sim 10^{-6}$. As demonstrated above, in our calculations, the condition $\int \Omega_{gw}(f)d\ln(f)\ll\Omega_\phi$ is always maintained, thus the ansatz neglecting the backreaction from GWs to axion dynamics can always hold.

The last but not the least, in the epoch $H\ll m_a$ and $\ddot{a}/a\ll m_a^2$, Eq.(\ref{eqphi}) has a solution $\phi(t)\approx \phi(t_i)(a/a_i)^{-3/2}\cos\left(m(t-t_i)\right)$, thus the energy density $\rho_{\phi}\propto a^{-3}$, which behaves like the dark matter. To avoid a large dark matter relic density, one way out is that $\phi$ could decay to other particles such as dark photons\cite{Machado:2018nqk}.  Meanwhile, for all BPs, the axion decay widths fulfill the condition $\Gamma\ll m_a $ to guarantee that the axion can generate GWs before its decay.

\section{ Regularization of ghost modes by higher order terms}
\label{sghost}

To study the tachyonic enhancement and narrow parameter resonance of GWs and to cure two issues of the model, we introduce a regularization procedure where a time-dependent cutoff is introduced. One may consider a time-independent cutoff to regularize ghost modes. For example, in Ref.\cite{Dyda:2012rj}, a constant cutoff scale $\Lambda_c$ on the physical wavenumber $k/a$ is introduced to avoid an infinite vacuum decay rate. By introducing a constant cutoff $k < \Lambda_c < 0.2 \, min(|m_{cs}(t)|)$, we found the total enhancement factor of GWs can not be larger than $20\%$, since the narrow parameter resonances are completely omitted. While those momentum modes $k > \Lambda_c$ do exist in the universe before the epoch $H(t^*)=m_a$, which are produced by other physical processes like inflation and phase transitions. How these modes are affected by the Chern-Simons term is physical. Since the coupling between these modes and the axion field is proportional to $\frac{k^2}{m_{cs}^2 a^2}$, which varies with time and is strong when $k$ is large and $m_{cs}\sim m_a$, it demonstrates that the linear expansion approximation is broken down. Some more appropriate treatments are needed. 

One may introduce high-order operators in the effective Lagrangian to regularize ghost modes. For example, we can consider the following higher order operators 
\begin{equation}\label{keyoprat}
c_{1}\partial^{\mu}\phi\partial^{\nu}\phi R_{\mu}^{\ \ \ \alpha}R_{\nu\alpha}+c_{2}\partial^{\mu}\phi\partial^{\nu}\phi R_{\mu\rho\nu\beta}R_{\alpha}^{\ \rho\alpha\beta}+c_{3}\partial^{\mu}\phi\partial^{\nu}\phi R_{\mu\rho\alpha\beta}R_{\nu}^{\ \rho\alpha\beta}	\,.
\end{equation}
Thus, the factor $D_A$ in the EOM of GWs given in Eq. (\ref{eqhA}) can be modified as $\tilde{D}_A$
\begin{equation}
	\begin{aligned}
		D_A\rightarrow \tilde{D}_A&=1- \frac{\lambda_Ak}{am_{cs}}+ \frac{b}{4}\left(\frac{k}{am_{cs}}\right)^2, A=R,L \label{horeg}
	\end{aligned}
\end{equation}
where $b=c_3$ when $c_1=c_2=0$. Thus if $b>1$, the EOM can be applied to all momentum modes and there is no ghost mode in the Lagrangian.

Generically speaking, to cure the ghost mode issue, we can resort to higher-order operators. For example, higher-order operators can include the following quadratic terms of graviton field terms (we take the traceless transverse gauge and assume that GWs travel in the direction of the $z$ axis)
\begin{equation}
(\partial\phi)^{2 m} \left  [  (\partial_z^{2n} h_{L}^*)  \, (\square h_{L}) + (\partial_z^{2n} h_{R}^*) \,\,(\square h_{R} ) \right ] \label{quadratict} \,,
\end{equation}
They can cure the ghost modes for any momentum modes $k$ and guarantee the Hamiltonian is bounded below. It should be pointed out that to have such quadratic terms, the Lorentz symmetry must be broken. Since $\theta$ is assumed to vary only with time, it is an obvious source of Lorentz symmetry breaking. 

When more high operators can be introduced in a specific fashion, the factor $D_A$ can even be modified as
$E_A$
\begin{equation}
	\begin{aligned}
		D_A\rightarrow E_A&= \exp( - \frac{\lambda_A k}{a m_{cs}} ) \label{expreg}\,.
	\end{aligned}
\end{equation}
Thus in this treatment, there are no ghost modes and all momentum modes can be treated on the same footing.

Although the ghost modes can be removed from the Lagrangian/Hamiltonian, it is found that high $k$ modes do not decouple either for the case $\tilde{D}_A$ or for the case $E_A$, which is one of the typical issues with high derivative theories \cite{Ostro,Woodard:2015zca,Aoki:2020gfv}. The enhancement of GWs depends upon the parameter $b$ for the case $\tilde{D}_A$ and the enhancement factor is larger when $k$ is larger for both cases, which can invalidate the ansatz of neglecting the backreaction of GWs to the evolution axion field.

To be consistent with the ansatz of neglecting the backreaction of GWs, a UV cutoff of momentum modes must be introduced. The effects of the backreaction of GWs to the axion field can be expressed as given below
\begin{equation}
	\begin{aligned}
		\ddot{\phi}&+3H\dot{\phi}+\frac{\partial V}{\partial \phi}=-\frac{\alpha}{4} R\tilde{R}\\
		R\tilde{R}&=4\int d^3k \left(\frac{k^3}{a^3}h_L^*(k)\dot{h}_L(k)-\frac{k}{a}\ddot{h}_L^*(k)\dot{h}_L(k)-\frac{k}{a}H\dot{h}_L^*(k)\dot{h}_L(k)-(L\Longleftrightarrow R)\right)
	\end{aligned}
\end{equation}
The $\tilde{R}R$ term on the right side of the EOM of $\phi$ makes a divergent contribution unless a reasonable cutoff scale is introduced. Thus, to make our ansatz hold, we use the hard cutoff $\Lambda_c$ and modify the factor $D_A$ into $\bar{D}_A$
\begin{equation}
	\begin{aligned}
		D_A\rightarrow \bar{D}_A&=1-\left(\frac{\lambda_Ak}{am_{cs}}-\frac{b}{4}\left(\frac{k}{am_{cs}}\right)^2\right)\Theta\left(1-\frac{k}{a\Lambda_{c}}\right), A=R,L\,.
	\end{aligned}
\end{equation}
By doing such a modification, we can cure the issues of ghost modes and avoid the backreaction effects.

\begin{figure}[htb]
	\centering
	\includegraphics[width=0.7\linewidth]{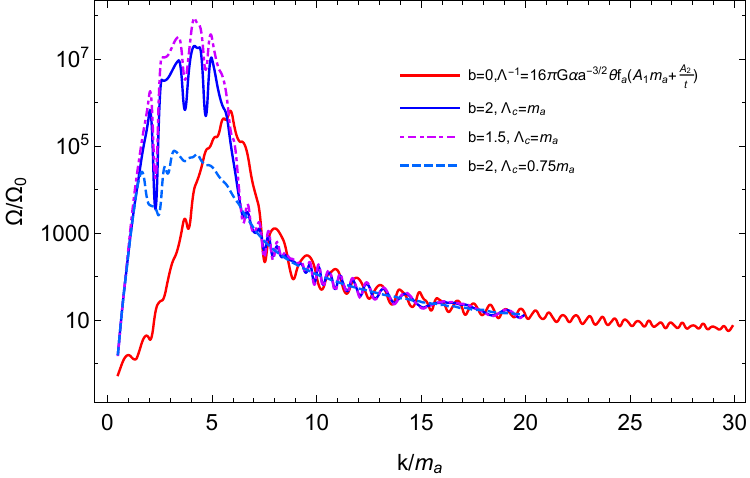}
	\caption{Magnification of energy density for two regularization procedures are compared with BP1 model parameters. The red line denotes the results of the procedure which simply turns off the CS term when $k/a> |m_{cs}| $, while the other lines denote the results of the procedure of higher order operators plus a constant cutoff. }
	\label{fig:plot33}
\end{figure}

The final results can depend upon the parameter $b$ and cutoffs. To compare, the magnification of energy density for two regularization procedures is shown in Fig.\ref{fig:plot33}. the red line is the results of the regularization method introduced in section \ref{s2} and other lines correspond to new regularization procedures by introducing higher order operators and hard cutoff scales. 

From Fig. \ref{fig:plot33}, it is observed that the regularization procedures can significantly affect the peak of magnification but for $k/m_a>8$ the result is almost the same for different regularization procedures. This is because, for these modes, the narrow parametric resonance happens at times when $|m_{cs}|$ evolves to a value much larger than $m_a/2$, the higher order operators have less effect on narrow parametric resonance.

It should be noteworthy that the enhancement factors and line shapes do depend upon regularization procedures and model parameters. For example, for the regularization procedure of high order operator, the peak modes situated at the region $1<k/m_a<7$ and enhancement factor $\Omega/\Omega_0$ at the peak region can be larger than $10^4\sim10^7$.


\section{discussion and summary}
\label{s4}
In this work, we propose a novel generation mechanism for GWs,  as well as an efficient way to detect axions/ALPs. The tachyonic instability and parametric resonance in the axion-graviton coupling is a less studied effect previously, and the axion-graviton coupling is much less constrained compared to the axion-photon coupling. The Chern-Simons axion-graviton coupling is an interesting modification to Einstein gravity, an extension of the effective field theory of Einstein-Hilbert action. This Chern-Simons gravity term can generate axion-photon coupling at the loop level with Planck suppression. The large coupling of axion-graviton can generate GWs in a radiation-dominated era, much later than inflation and preheating time which people usually explored.

To summarize this work, we would like to emphasize that the peak frequency of the generated GWs is red-shifted today and can be detected by different GW detection methods, from $10^{-9}$Hz up to $10^{-2}$Hz. We have also numerically studied the shape of the GW spectrum, in the ansatz neglecting the backreaction of GWs to the axion field. The upper bounds of this power spectrum are mostly from observation, as we discussed earlier. The parameters we used in the figure are well below the bounds from $\Delta N_{eff}$ in the radiation era, so this mechanism can be a source for large dark radiation.

\begin{acknowledgments}
S. Sun is supported by the National Natural Science Foundation of China (Nos. 12105013).
Q. Yan is supported by the Natural Science Foundation of China under Grants No. 11875260 and No. 12275143.
Z. Zhao has been partially supported by a China and Germany Postdoctoral Exchange Program between
the Office of China Postdoctoral Council (OCPC) and DESY.
\end{acknowledgments}

\bibliographystyle{apsrev4-1}
\bibliography{myrefs.bib}
\end{document}